\documentclass[12pt]{JHEP3}

 \preprint{  }

\usepackage{epsfig,multicol}
\usepackage{amsmath}
\usepackage{graphics}
\usepackage{graphicx}
\usepackage{dcolumn}
\usepackage{amssymb}
\usepackage{dcolumn}

\def \be {\begin{equation}}
\def \ee {\end{equation}}
\def \bea {\begin{eqnarray}}
\def \eea {\end{eqnarray}}

\title{\bf  Holographic entanglement entropy in metal/superconductor phase transition with exponential nonlinear electrodynamics}

\author{Weiping Yao \\ Department of Physics and Electronic Science, Liupanshui Normal University, Liupanshui, Guizhou 553004, P. R. China.}

\author{Jiliang Jing\footnote{Corresponding author, Email: jljing@hunnu.edu.cn}
\\ Department of Physics, Key Laboratory of Low Dimensional \\ Quantum Structures and Quantum Control of Ministry of Education, and Synergetic Innovation Center for Quantum Effects and Applications, Hunan Normal University, Changsha, Hunan 410081, P. R. China.}

\abstract{
We study the holographic entanglement entropy in metal/superconductor phase transition
with exponential nonlinear electrodynamics (ENE) in four and five dimensional spacetimes.
We find that the holographic entanglement entropy is powerful tool in studying the properties of the holographic phase transition. For the operator $\langle\mathcal{O}_{+}\rangle$,  we observe that the entanglement entropy in 4-dimensional spacetime decreases in metal phase but changes non-monotonously in superconducting phase with the increase of the ENE parameter.  Interestingly, the change of the entanglement entropy in 5-dimensional spacetime for both the phases is monotonous as the ENE factor alters. For the operator $\langle\mathcal{O}_{-}\rangle$, we show that the behavior of entanglement entropy in four and five dimensions changes monotonously for both phases as one tunes the strength of the ENE. Furthermore,  for both the operators the entanglement entropy in four or five dimensional black hole increases with the increase of the width of the region.
}


\begin{document}

\section{Introduction}
The anti-de Sitter/ conformal field theory (AdS/CFT) correspondence states that
a weak coupling gravity theory in a d-dimensional AdS spacetime can be related to
a strong CFT on the (d-1)-dimensional boundary \cite{J.Maldacena1998, EWritten1998, S.S.Gubser1998}. Based on this novel idea, the simplest example of the duality between the
superconductor and gravity theory was first constructed in \cite{S.A.Hatnall2008}.
It was suggested that the Reissner-Nordstrom-AdS (RN-AdS) black hole becomes unstable and finally  the scalar hair brings spontaneous $U(1)$-gauge symmetry breaking as one tunes the temperature for black hole. And the RN-AdS and hair black holes correspond to a second order phase transition from normal phase to superconducting phase. Since this prominent connection between condensed matter and gravitational physics proved us a elegant approach to study the strongly coupled condensed matter systems, a lot of the holographic superconductor
models have been widely studied in Refs.\cite{maeda, R.Gregory 2009, Q.Y.Pan2009, CaiNie, Herzog-2010, Franco, Jing2010, jingpanl, Sunandan Gangopadhyay2012, Qiyuan Pan2012, Zixu Zhao2012, Davood Momeni2013, Weiping Yao2013, Sunandan Gangopadhyay2014, Yan Peng2015,
Lukasz Nakonieczny2014,  P Chaturvedi2015, Yunqi Liu2015}. On the other hand, the entanglement entropy is expect to be an effective measure of how a given
quantum system is strongly correlated. In the spirit of AdS/CFT correspondence, Ryu and Takayanagi \cite{Ryu:2006bv, Ryu:2006ef} have provided a geometric proposal to compute the entanglement entropy
of CFTs from the minimal area surface in gravity side. Recently, this elegant approach
has been applied extensively to study various properties of holographic superconductors
\cite{Nishioka:2009un,Albash:2011nq,Myers:2012ed,deBoer:2011wk,Hung:2011xb,Nishioka:2006gr,
Klebanov:2007ws,Pakman:2008ui} and it has been proven as a good probe to the critical phase transition point and the order of phase transition \cite{Ogawa:2011fw, Tameem Albash2012, Xi Dong2013, Cai2012, Li-Fang Li2013, Cai1203, Kuang2014, Y Peng2015, Peng2015}.

With these considerations in mind, we have studied the holographic entanglement entropy in the phase transition for the Born-Infeld electrodynamics in Refs \cite{yao2014, yao2014a}. Interestingly enough, the
Exponential nonlinear electrodynamics (ENE), which was introduced by Hendi to obtain the new charged BTZ black hole solutions in the Einstein nonlinear
electromagnetic field, is a new Born-Infeld-like nonlinear electrodynamics\cite{Hendi2012,Hendi2013}.
The Lagrangian density for the ENE theory is
\begin{equation}\label{L:ENE}
L_{ENE}=\frac{1}{4b^2}\bigg[e^{-b^2 F^2}-1\bigg],
\end{equation}
$F^2=F^{\mu\nu}F_{\mu\nu}$, where $F^{\mu\nu}$ is the electromagnetic field tensor.
b is the Exponential coupling parameter and this Lagrangian will reduce to the Maxwell case as the factor b approaches zero. Compared to the Born-Infeld nonlinear electrodynamics (BINE) \cite{M.Born1934, B.Hoffmann1935, W.Heisenberg1936, Oliveira1994, G.W.Gibbons1995},
the ENE displays different effect on the on the electric potential and temperature for
the same parameters and its singularity is much weaker than the Einstein-Maxwell theory \cite{Hendi2014,A. Sheykhi, A. Sheykhi2014, Hendi2015}. Therefore it is of great interesting to see the phase transition
in the frame of the ENE theory, in particular, to see how the entanglement entropy changes as the strength of the ENE changes.

In the present paper, we would like to investigate the behavior of entanglement entropy
in the metal/superconductor phase transition with Exponential nonlinear electrodynamics
to examine whether the holographic entanglement entropy is still useful in describing properties of the phase transition system. In the 4-dimensional spacetime, we observe that
the entanglement entropy decreases in the metal phase for both operators as the
ENE factor becomes larger. In the superconducting phase, however, with the increase of the ENE factor, the entanglement entropy of the operator $\langle\mathcal{O}_{-}\rangle$ increases monotonously but the entanglement entropy first increases and then decreases continuously for the operator $\langle\mathcal{O}_{+}\rangle$. Considering the higher dimension, interestingly enough, we note that with the increase of the ENE parameter the value of the entanglement entropy increases monotonously in 5-dimensional spacetime  for the two operators in the superconducting phase. While the entanglement entropy in the normal phase decreases as the ENE factor becomes larger.
 We also note that the slope of entanglement entropy in four and five dimensional spacetimes
presents a discontinuous change at a critical temperatures which implies a
significant reorganization of the degrees of freedom of the system due to
the process of phase transition. Furthermore, the jump of the slope of the entanglement entropy
 can be regarded as the signature of the second order phase transition.

The structure of this work is as follows. In section II, we will introduce the holographic
superconductor model with ENE theory and study the properties of
holographic phase transition by examining in detail the scalar operator.
In section III, we will investigate the holographic entanglement entropy of the
phase transition system in four and five dimensional spacetimes, respectively.
In section V, we will summarize our results.

\section{Holographic Superconducting Models with Exponential electromagnetic field}
\subsection{equations of motion and boundary conditions}
The action for the exponential electromagnetic field coupling to a charged scalar field in Einstein gravity with a cosmological constant is
\begin{eqnarray}\label{action}
S&=&\int d^{d}x\sqrt{-g}\Big\{\frac{1}{16\pi G_{d}}\left[R+\frac{(d-1)(d-2)}{L^{2}}\right]
\nonumber \\ &&+\left[-|\nabla\Psi-iq A\Psi|^2-m^2|\Psi|^2\right]+L_{ENE}\Big\},
\end{eqnarray}
where $g$ is the determinant of the metric,
$L$ is the radius of AdS spacetime, $\psi$ stands for the complex scalar field,
A is the gauge field and $F = d A $ is the strength of a U(1) gauge field,
$q$ is the coupling parameter between the gauge field and the complex scalar
field,  $m$ is the mass of the scalar field, and  $L_{ENE}=\frac{1}{4b^2}[e^{-b^2 F^2}-1]$ is
Lagrangian density for the ENE theory. In the limit $b\rightarrow0$,
the ENE field will reduce to the Maxwell field.

To construct a superconductor dual to  AdS black hole configuration,
we take the coordinates for the planar black hole with full backreaction and the matter
fields in the forms
\begin{eqnarray}\label{BH metric}
&&ds^2=-f(r)e^{-\chi(r)}dt^{2}+\frac{dr^2}{f(r)}+r^{2}h_{ij}dx^{i}dx^{j}, \\
&&A=\phi(r)dt,~~\psi=\psi(r).
\end{eqnarray}
The Hawking temperature of this black hole is
\begin{eqnarray}\label{Hawking temperature}
T_{H}=\frac{f^{\prime}(r_{+})e^{-\chi(r_{+})/2}}{4\pi},
\end{eqnarray}
where $r_+$ is the horizon of the black hole.

From the action and the metric, we can obtain the corresponding independent
equations of motion as follows
\begin{eqnarray}\label{psi}
&& \psi^{\prime\prime}+\left(\frac{d-2}{r}-\frac{\chi^{\prime}}{2}+
\frac{f^\prime}{f}\right)\psi^\prime+\frac{1}{f}\left(\frac{q^{2}e^{\chi}\phi^2}{f}-m^2\right)
\psi=0,
\\ \label{phi}
&& (1+4\beta^2e^\chi \phi'^2)\phi^{\prime\prime}+\left(\frac{d-2}{r}+\frac{\chi^{\prime}}{2}\right)\phi^\prime+
2\beta^2e^{\chi}\chi'\phi'^3-
\frac{2q^{2}\psi^{2}e^{-2\beta^2e^\chi \phi'^2}}{f}\phi=0,
\\
&& \chi^{\prime}+\frac{2r}{d-2}\left(\psi^{\prime
2}+\frac{q^{2}e^{\chi}\phi^{2}\psi^{2}}{f^{2}}\right)=0,\label{chi}\\
&& f^{\prime}-\left(\frac{(d-1)r}{L}-\frac{(d-3)f}{r}\right)+\frac{r}{d-2}
\Big[m^{2}\psi^{2}
+f\left(\psi^{\prime
2}+\frac{q^{2}e^{\chi}\phi^{2}\psi^{2}}{f^{2}}\right)\nonumber \\ &&~~~~+
\frac{1-e^{2\beta^2e^\chi \phi'^2}(1-4\beta^2e^\chi \phi'^2)}{4\beta^3}\Big]=0,
\end{eqnarray}
where the prime denotes the derivative with respect to $r$, and $16\pi G=1$ was used.
In order to solve the equations of this complex system,
we have to specify the boundary conditions.
At the horizon $r_+$, the regularity condition of the boundary conditions behave as
\begin{eqnarray}
\phi(r_+)=0,\qquad f(r_+)=0.
\end{eqnarray}
And at the asymptotic AdS boundary ($r\rightarrow\infty$), the
asymptotic behaviors of the solutions are
\begin{eqnarray}
\chi\rightarrow0\,,\hspace{0.5cm}
f\sim r^2\,,\hspace{0.5cm}
\phi\sim\mu-\frac{\rho}{r^{d-3}}\,,\hspace{0.5cm}
\psi\sim\frac{\psi_{-}}{r^{\Delta_{-}}}+\frac{\psi_{+}}{r^{\Delta_{+}}}\,,
\label{infinity}
\end{eqnarray}
where $\mu$ and $\rho$ are interpreted as the chemical potential and
charge density in the dual field theory, and the exponent $\Delta_\pm$ is defined by
$((d-1)\pm\sqrt{(d-1)^2+4m^{2}})/2$ for d-dimensional spacetime. Notice that, provided $\Delta_{-}$ is larger than the unitarity
bound, both $\psi_{-}$ and $\psi_{+}$ can be normalizable. According to the AdS/CFT correspondence, they correspond to the vacuum expectation values $\psi_{-}=\langle\mathcal{O}_{-}\rangle$, $\psi_{+}=\langle\mathcal{O}_{+}\rangle$ of an operator $\mathcal{O}$ dual to the scalar field
\cite{HartnollPRL101}. In the following calculation, we impose boundary condition that either  $\psi_{-}$ or $\psi_{+}$ vanishes.

From the equations of motion for the system, we can get the useful scaling symmetries
in the forms
\begin{eqnarray}
&& r\rightarrow \alpha r,\qquad (x,y,t)\rightarrow(x,y,t)/\alpha,\qquad\phi\rightarrow \alpha\phi,\qquad f\rightarrow\alpha^2f,\label{scaling:r}\\
&& L\rightarrow\alpha L,\qquad r\rightarrow \alpha r,\qquad t\rightarrow \alpha t,\qquad q\rightarrow\alpha^{-1} q, \label{scaling:L}\\
&& e^{\chi}\rightarrow\alpha^2e^{\chi},\qquad\phi\rightarrow \alpha^{-1}\phi,\qquad t\rightarrow t\alpha. \label{scaling3}
\end{eqnarray}
Using the scaling symmetries (\ref{scaling:r}), we can take $r_+=1$ and
the symmetries (\ref{scaling:L}) allow us to set the $L=1$.
Then, the following quantities can be rescaled as
\begin{equation}\label{scaling Q}
\mu\rightarrow\alpha\mu,\qquad
\rho\rightarrow\alpha^{d-2}\rho, \qquad
\langle\mathcal{O}_{-}\rangle\rightarrow\alpha^{\Delta_-}\langle\mathcal{O}_{-}\rangle,\qquad
\langle\mathcal{O}_{+}\rangle\rightarrow\alpha^{\Delta_+}\langle\mathcal{O}_{+}\rangle.
\end{equation}

\subsection{Physical properties of phase transition}
We now study the physical properties of phase transition in this
model through the behaviors of the scalar condensation.
In order to obtain the solutions in metal phase, we set the
function $\psi(r)=0$ and the solution of the system is just the
black hole with exponential form of nonlinear electrodynamics.
However,
the equations are
nonlinear and coupled to each other in the superconducting phase where $\psi(r)\neq 0$. Therefore, we solve these equations by using the numerical shooting method. Introducing a new variable $z=r_{+}/r$,  then the region $r_+<r<\infty$
correspondences to $1>z>0$. In the following, we will work in the phase transition
of the complex model and study the effects of the ENE
parameter $b$ on the condensation of the scalar operators in d-dimensional spacetime. And
the useful dimensionless quantities are
\begin{eqnarray}\label{dimensionless quantities}
T/\rho^{\frac{1}{d-2}}, \qquad
\langle\mathcal{O}_{-}\rangle^{\frac{1}{\Delta_-}}/ \rho^{\frac{1}{d-2}},\qquad
\langle\mathcal{O}_{+}\rangle^{\frac{1}{\Delta_+}}/ \rho^{\frac{1}{d-2}}.
\end{eqnarray}

Firstly, we concretely investigate the phase transition for ENE field with full backreaction in the 4-dimensional AdS black hole spacetime. Considering Breitenlohner-Freedman bound\cite{Breitenlohner 1982}, we here set $m^2=-2$ and $q=1$. The scalar operator as a function of temperature with different
values of ENE parameter is shown in Fig.~\ref{condesation d4}.
In the left panel,
it is shown that the operator $\langle\mathcal{O}_{+}\rangle$
emerges at critical temperature $T_c$, which implies
the phase transition appears. The critical behavior near $T_c$ is found
to be $\langle\mathcal{O}_{+}\rangle \varpropto (1-T/T_c)^{\frac{1}{2}}$, which represents
the transition is second order. It should be noted that the
value of the critical temperature $T_c$ becomes smaller as the
ENE factor $b$ increases, which means that the ENE
correction to the usual Maxwell field makes the scalar hair harder
to form in the full-backreaction model.
Furthermore, for a given $b$ ,the critical temperature $T_c$ of the ENE system is smaller than that of the BINE system which was studied in Ref.\cite{yao2014a}. That is to say, the scalar hair is more difficult to be develop  in a model with ENE than that with BIEN.
For the operator $\langle\mathcal{O}_{-}\rangle$ (right plot), we also find that
there is a phase transition at the critical temperature $T_c$. And the critical temperature $T_c$ decreases with the increase of the parameter $b$. As the factor $b$ approaches to zero,
our result is consistent with the result in Ref. \cite{HartnollPRL101}.

\FIGURE{
\includegraphics[scale=0.85]{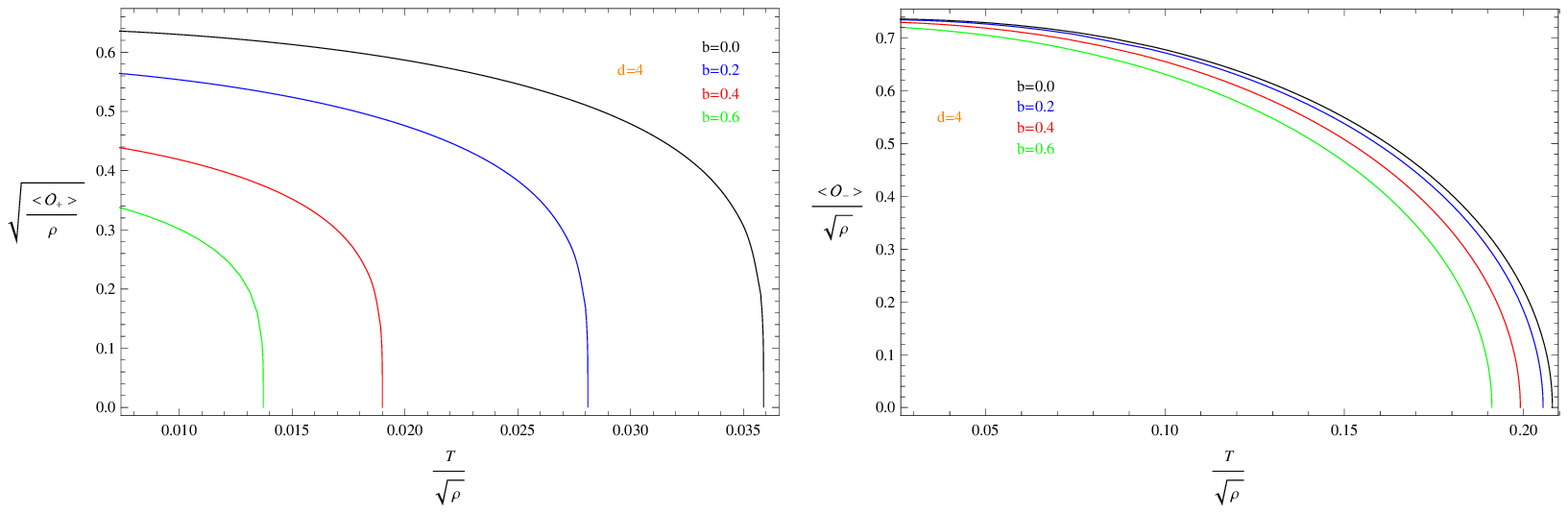}
\caption{ (Color online.) The operators $\langle\mathcal{O}_{+}\rangle$ (left plot) and $\langle\mathcal{O}_{-}\rangle$
(right plot) versus temperature after condensation with different
values of ENE parameter in 4-dimensional spacetime. The four lines from top to bottom
correspond to $b=0$ (black), $b=0.2$ (blue), $b=0.4$ (red), and
$b=0.6$ (green) respectively.} \label{condesation d4}}

Then, it is interesting to study the behavior of the condensation in higher dimensional spacetimes. Concretely, we set $d=5, q = 2, m^2 = \sqrt{-15/4}$. In Fig. \ref{condesation d5}, we plot the behaviors of condensation with
the changes of the temperature and the ENE parameter in 5-dimensional spacetime.
\FIGURE{
\includegraphics[scale=0.85]{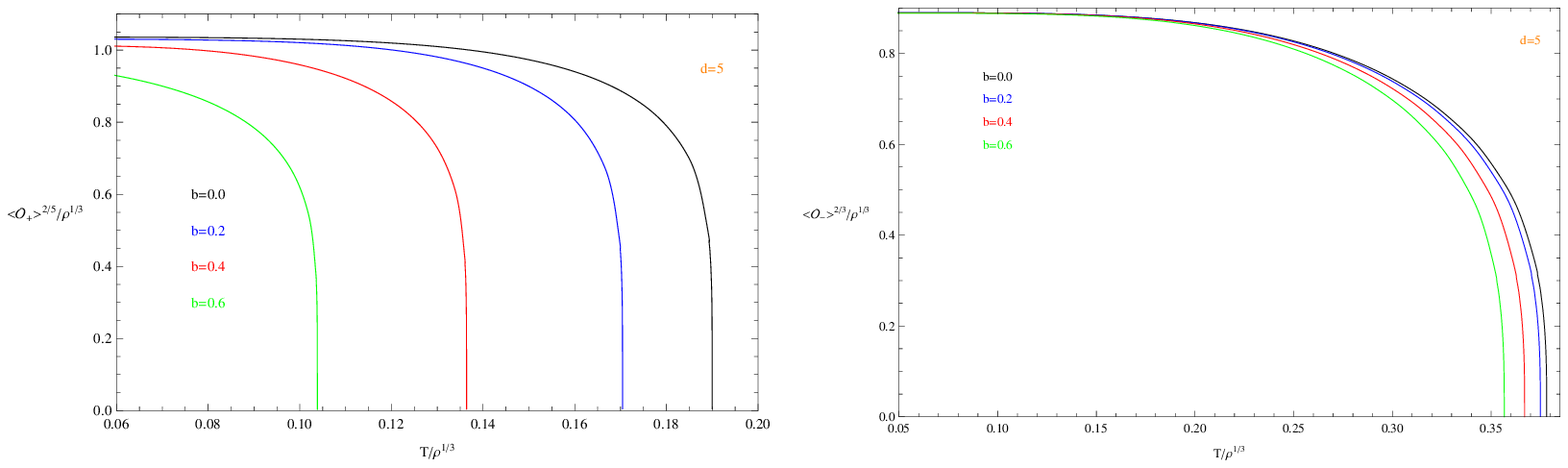}
\caption{ (Color online.) The operators $\langle \mathcal{O}_{+}\rangle$ (left plot) and $\langle \mathcal{O}_{-}\rangle$
(right plot) as a function of the temperature after condensation with different
values of ENE parameter in 5-dimensional spacetime. The four lines from top to bottom
correspond to $b=0$ (black), $b=0.2$ (blue), $b=0.4$ (red), and
$b=0.6$ (green) respectively.} \label{condesation d5}}
From the left plot of Fig.~\ref{condesation d5}, we observe that when the temperature
decreases to be lower than the critical value $T_c$, the
condensation of the operator $\langle\mathcal{O}_{+}\rangle$ appears and this tells
us that the scalar hair forms. Interestingly, with the
increase of the ENE factor $b$, the value of the critical temperature $T_c$ decreases.
This means that the ENE correction to the usual Maxwell field hinders the formation of
the scalar condensation in this system.
In the right-hand panel, we find that the behaviors of condensate of the operator $\langle\mathcal{O}_{-}\rangle$ with the changes of the temperature
and the ENE parameter are similar to the case for the operator $\langle \mathcal{O}_{+}\rangle$.

\section{Holographic entanglement entropy of the transition system }
In this section, we will concentrate on the holographic entanglement entropy
of the holographic model. A simple and elegant proposal to entanglement entropy of the boundary field theory in the context of gauge/gravity duality has been presented in
Refs.\cite{Ryu:2006bv, Ryu:2006ef}. It is explicitly given by
the formula
\begin{equation}\label{law}
S_A=\frac{\rm Area(\gamma_\mathcal{A})}{4G_N},
\end{equation}
where $S_A$ is the entanglement
entropy for the subsystem A which can be chosen arbitrarily,
$\gamma_A$ is the minimal area surface in the bulk which ends on the boundary of $A$,
$G_N$ is Newton constant in the Einstein gravity on the AdS space.
Here, we consider the entanglement entropy for a strip geometry in d-dimensional spacetime and study the behavior of holographic entanglement entropy in this transition system.

\subsection{Holographic entanglement entropy in 4-dimensional spacetime}
We first explore the influence of the ENE electrodynamics on the entanglement entropy by the holographic method in 4-dimensional spacetime. Considering the entanglement entropy for a strip geometry corresponding to a subsystem $A$
which is described by$-\frac{\ell}{2}\leq x \leq \frac{\ell}{2}\ $ and $-\frac{R}{2}<y<\frac{R}{2}~(R\rightarrow\infty)$, where $\ell$ is defined as the size of region $A$. Then, the entanglement entropy in the strip geometry can be obtained in the following
\begin{eqnarray}\label{EEntropyd4}
S_A=\frac{R}{2G_4}\int^{z_{*}}_{\varepsilon}dz\frac{z_{*}^{2}}{z^{2}}\frac{1}{\sqrt{(z^{4}_{*}-z^{4})z^{2}f(z)}}
=\frac{R}{2G_4}\left(\frac{1}{\varepsilon^2}+s\right),
\end{eqnarray}
and the width $\ell$ is
\begin{eqnarray}\label{Length}
\frac{\ell}{2}=\int^{z_{*}}_{\varepsilon}dz\frac{z^{2}}{\sqrt{(z^{4}_{*}-z^{4})z^{2}f(z)}},
\end{eqnarray}
where $z=\frac{1}{r}$ and $z_{*}$ satisfies the condition $\frac{dz}{dx}|_{z_{*}}=0$.
In Eq. (\ref{EEntropyd4}), the first term $\frac{1}{\epsilon^2}$ is UV cutoff and
represents the °∞area law°± \cite{Ryu:2006bv, Ryu:2006ef}. The second term s is finite term
and thus is physically important.
According to the discussion in the part A of
Section II, the above physical quantities $s$ and $\ell$
under the Eq. (\ref{scaling:r}) can be rescaled as
$s\rightarrow\alpha s \ \ \ , \ell\rightarrow\alpha^{-}\ell$.
Consequently, we can use these scale invariants in the following calculation
\begin{equation}\label{invariant:s}
s/\sqrt{\rho} \ \ \ , \ell \sqrt{\rho}.
\end{equation}

The behavior of the entanglement entropy of the operator $\langle O_{+}\rangle$ as a function of temperature $T$ and the nonlinear factor $b$
with $\ell\sqrt{\rho}=1$ is presented in Fig. \ref{d4gzbst}.
The vertical dashed
lines represent the critical temperature of the phase transition for
each b. The dot-dashed lines
are from the normal phase and the solid ones are from the
superconducting cases.
It is found that the entanglement entropy as a function of temperate has a discontinuous slop at the critical temperature $T_c$. Which indicates
that some kind of new degrees of freedom like the Cooper pairs
would appear in new phase. And this discontinuity corresponds to second order
phase transition in the holographic model with ENE.
Therefore, the entanglement entropy is useful in disclosing
the holographic pase transition, and that its behavior can indicate not only the appearance,
but also the order of the phase transition.

Moreover, the entanglement entropy after
condensate is lower than the one in the metal phase and decreases
monotonously as temperature decreases. It means that the condensate
turns on at the critical temperature and the formation of Cooper pairs
make the degrees of freedom decrease in the hair phase.
For a given $T$, the value of the entanglement entropy
becomes smaller as the ENE factor $b$ increases in the normal phase. In the superconducting phase, however, the entanglement entropy first increases (left plot)and then decreases monotonously (right plot) as the parameter b becomes larger.
\FIGURE{
\includegraphics[scale=0.85]{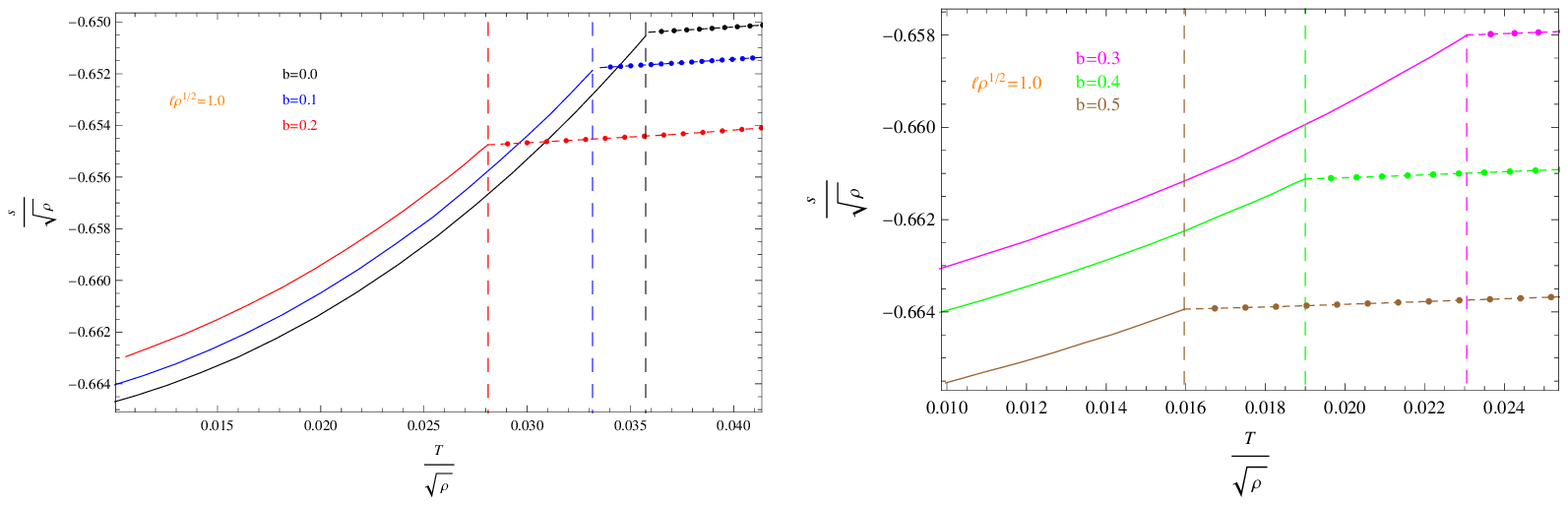}
\caption{\label{d4gzbst}(Color online.) The entanglement entropy of the
operator $\langle O_{+}\rangle$ as a function of the temperature $T$ and the
ENE factor $b$ for fixed belt width.
In the left plot, the three lines from bottom
to top correspond to $b=0$ (black), $b=0.1$ (blue) and $b=0.2$
(red), but in the right one are for $b=0.3$ (magenta), $b=0.4$
(green), $b=0.5$ (brown), respectively.}}

For the operator $\langle O_{-}\rangle$, interestingly, the dependence of the entanglement entropy on the factor $b$ is monotonic in the hair phase which is displayed in Fig. \ref{d4gfbst}. It also can
be seen from the picture that the value of the entanglement entropy in metal phase decreases with the increase of the parameter b. Furthermore, the value of the critical temperature $T_c$ decreases as we choose a lager b. Comparing with the results in Fig. \ref{condesation d4},
we mention that the entanglement entropy is more sensitive to the change of the nonlinear parameter b than the scalar operator, which provide us a better approach to detect the effects of exponential nonlinear electrodynamics.

\FIGURE{
\includegraphics[scale=0.6]{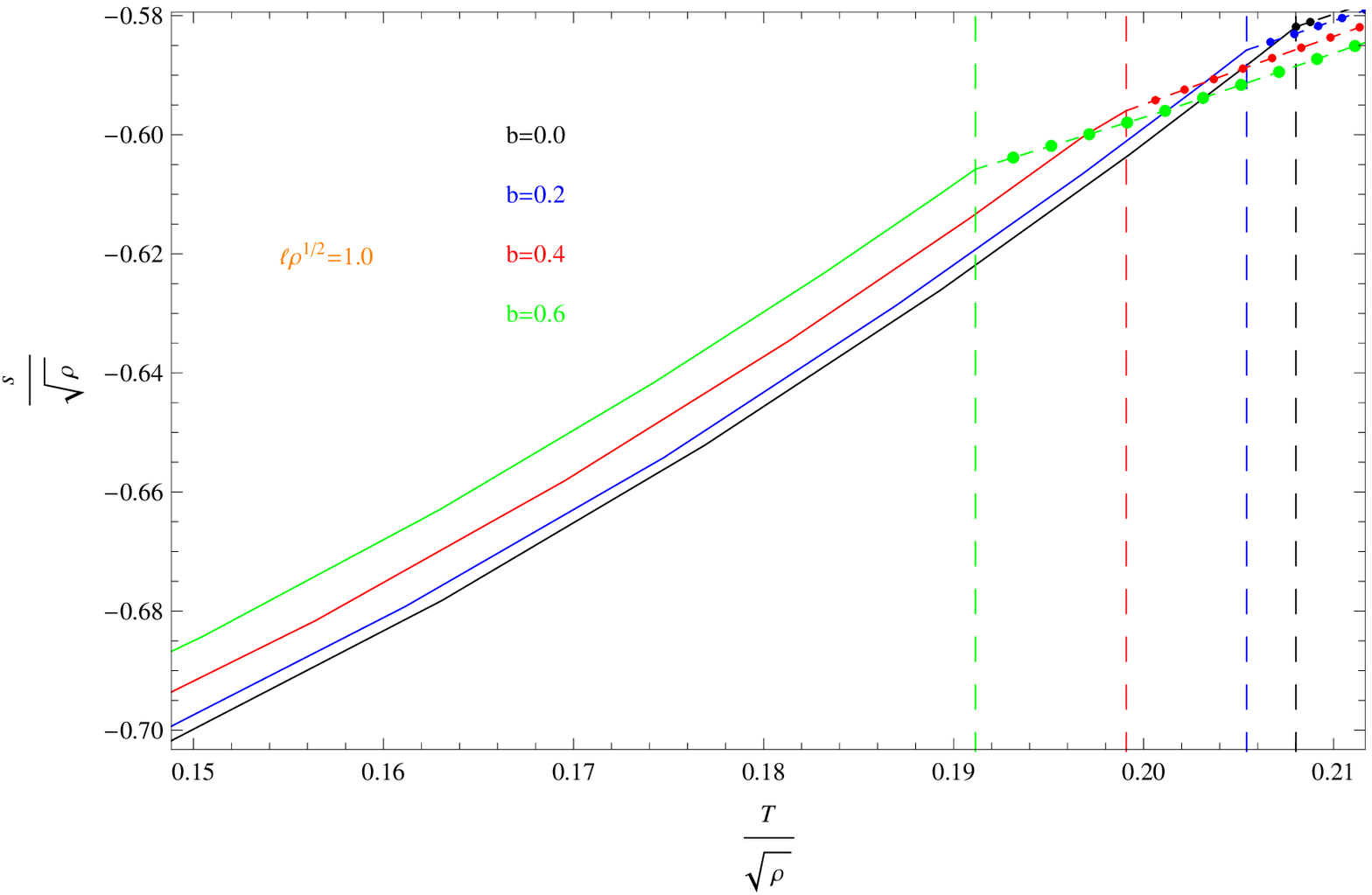}
\caption{\label{d4gfbst}(Color online.) The entanglement entropy of the
operator $\langle O_{-}\rangle$ as a function of the temperature $T$ and the ENE
 factor $b$ for fixed belt width. The vertical dotted lines represent the critical temperature of the phase transition for various b.
The dot-dashed line is for the metal phase and the solid
one is for the superconductor solutions
 The four lines
from bottom to top are for $b=0$ (black), $b=0.2$ (blue), $b=0.4$
(red), and $b=0.6$ (green) respectively.}}

Next, for the sake of studying the effect of strip width $\ell$ on the entanglement entropy in the superconducting phase, we plot in Figs. \ref{d4gzLst} and \ref{d4gfLst} that the behavior of the entanglement entropy as a function of the parameter $T$ for
different widths $\ell$ with $b=0.4$ for both operator $\langle O_{+}\rangle$ and operator $\langle O_{-}\rangle$.
From the Fig. \ref{d4gzLst}, we find that the entanglement entropy increases with the increase of belt width $\ell$. For the operator $\langle O_{-}\rangle$,  the entanglement entropy becomes
bigger with the increase of the factor $b$ which is displayed  the Fig. \ref{d4gfLst}.
\FIGURE{
\includegraphics[scale=0.85]{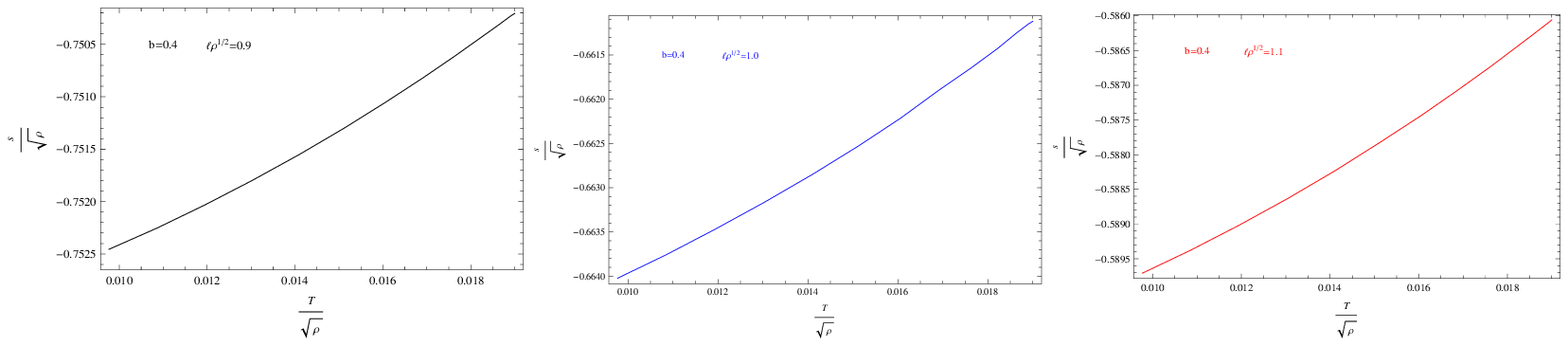}
\caption{\label{d4gzLst}(Color online.) The entanglement entropy of the
operator $\langle O_{+}\rangle$ is depicted as a function of T for
different widths $\ell$ with $b=0.4$. The left plot
(red) is for $\ell\sqrt{\rho}=1.1$, the middle one (blue) for
$\ell\sqrt{\rho}=1.0$, and the right one (black) for
$\ell\sqrt{\rho}=0.9$.}}
\FIGURE{
\includegraphics[scale=0.7]{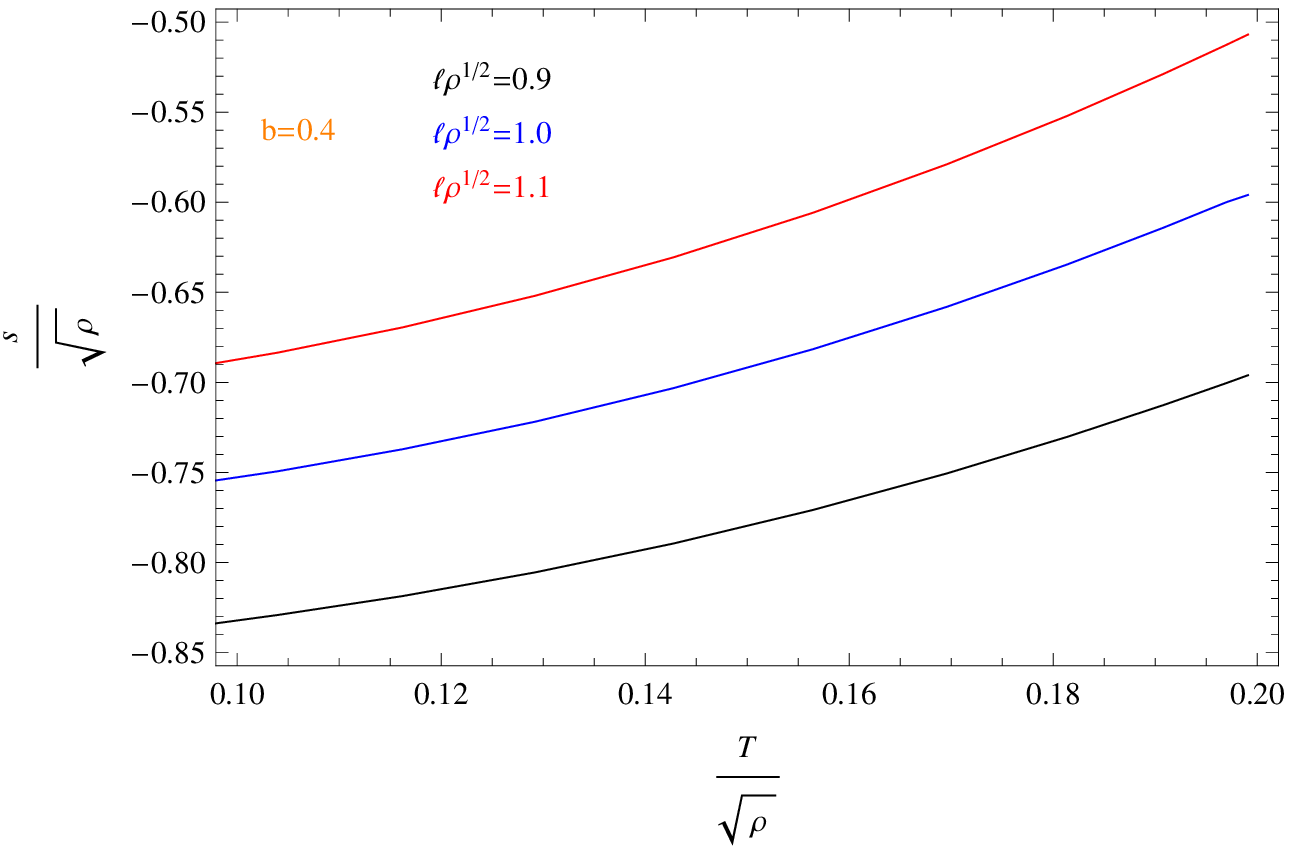}
\caption{\label{d4gfLst}(Color online.) The entanglement entropy of the
operator $\langle O_{-}\rangle$ is depicted as a function of T for
different widths $\ell$ with $b=0.4$. The three lines
from bottom to top are for $\ell\sqrt{\rho}=0.9$(black),
$\ell\sqrt{\rho}=1.0$ (blue), $\ell\sqrt{\rho}=1.1$ (red) respectively.}}

\subsection{Holographic entanglement entropy in 5-dimensional spacetime}

Considering the higher dimensional spacetime, we focus here on studying the
holographic entanglement entropy of the holographic model in 5-dimensional spacetime.
The entanglement entropy for a strip shape corresponding to a subsystem
$A$ is defined by a finite width $\ell$ along the x direction and infinitely extending in y and z directions. Thus, the geometry of the subsystem $A$ is described by
\begin{equation}\label{embed}
t=0,\ \ x=x(r),\ \ -\frac{R}{2}<y<\frac{R}{2}\ (R\rightarrow\infty),\ \ -\frac{W}{2} \leq z\leq \frac{W}{2}(W\rightarrow\infty).
\end{equation}
For the x direction, we assume that the holographic surface $\gamma_A$ starts from $r=\frac{1}{\epsilon}$ at $x=\frac{\ell}{2}$, extends into the bulk until it reaches $r=r_*$, then returns back to the AdS boundary $r=\frac{1}{\epsilon}$ at $x=-\frac{\ell}{2}$.
The induced metric on the hypersurface $A$ is
\begin{equation}
ds^2 =h_{ij}dx^i dx^j=\left (\frac{1}{f(r)}+r^2\left
(\frac{dx}{dr}\right )^2\right )dr^2+r^2 dy^2+r^2dz^2.
\end{equation}

According to the Eq. (\ref{law}), the entanglement entropy in the strip geometry can be expressed as
\begin{equation}\label{HEE}
S_\mathcal{A}[x]=\frac{RW}{2G_5}\int_{r_*}^{\frac{1}{\epsilon}}
r^2\sqrt{\frac{1}{f(r)}+r^2(dx/dr)^2}dr,
\end{equation}
where $r=\frac{1}{\epsilon}$ is the UV cutoff. Noting that the above expression
be treated as the Lagrangian with x direction thought of as time.
Therefore, we can obtain a constant of motion from Eq. (\ref{HEE})
\begin{equation}\label{d5minimal}
\frac{r^4(dx/dr)\sqrt{f(r)}}{\sqrt{1+r^2f(r)(dx/dr)^2}}=r_*^3 ,
\end{equation}
where we demand that the surface is smooth at $r=r_*$ i.e.
$dx/dr|_{r=r_*}=0$. By using the variable $z=1/r$ ,
the width $\ell$ in terms of $z$ can be written as
\begin{eqnarray}\label{d5Length}
\frac{\ell}{2}=\int^{z_{*}}_{\varepsilon}dz\frac{z^{3}}{\sqrt{(z^{6}_{*}-z^{6})z^{2}f(z)}},
\end{eqnarray}
and the holographic entanglement entropy in the $z-$coordinate is
\begin{eqnarray}\label{d5HEE}
S_A=\frac{R W}{2G_5}\int^{z_{*}}_{\varepsilon}dz\frac{z_{*}^{3}}{z^{2}}
\frac{1}{\sqrt{(z^{6}_{*}-z^{6})z^{2}f(z)}}
=\frac{R W}{4G_5}\left(\frac{1}{\varepsilon^2}+s\right).
\end{eqnarray}
the first term $1/\varepsilon^2$ is divergent as $\varepsilon\rightarrow0$.
The second term is independent of the cutoff and is finite, so it is physical important.
Now, we can calculate the entanglement entropy of the operators $\langle O_{-}\rangle$ and $\langle O_{+}\rangle$
and explore the effects of the temperature
$T$, the ENE factor $b$ and the belt width $\ell$ on the entanglement entropy
respectively.

The behavior of the entanglement entropy for the various factors is show in Fig. \ref{d5gzbst}
in the dimensionless quantities $s/ \rho^{\frac{1}{3}},\  \ell \rho^{\frac{1}{3}}, \  T/\rho^{\frac{1}{3}}$.
The vertical dotted lines represent the critical temperature of the phase transition for the different value of the ENE factor.
The dot-dashed line is for the normal phase and the solid one is for the superconductor solutions.
For both operators, the behavior of the entanglement entropy with respect of the
temperature is similar to the case of the 4-dimensional spacetime which is studied in section III.
Which means the entanglement entropy is also a good probe to
the holographic phase transition in higher dimensional spacetime of this model.
When the factor $T$ is fixed, for two operators we find that the entanglement entropy decreases (or increases) with the increase of the ENE factor b in the normal (or hair) phase. Furthermore, it is of interesting to note that
the slope of the curve decreases as the temperature is lowered in superconductor situation.
And the influence of the ENE parameter b on the entanglement entropy is getting smaller as we choose a low temperature.
\FIGURE{
\includegraphics[scale=0.85]{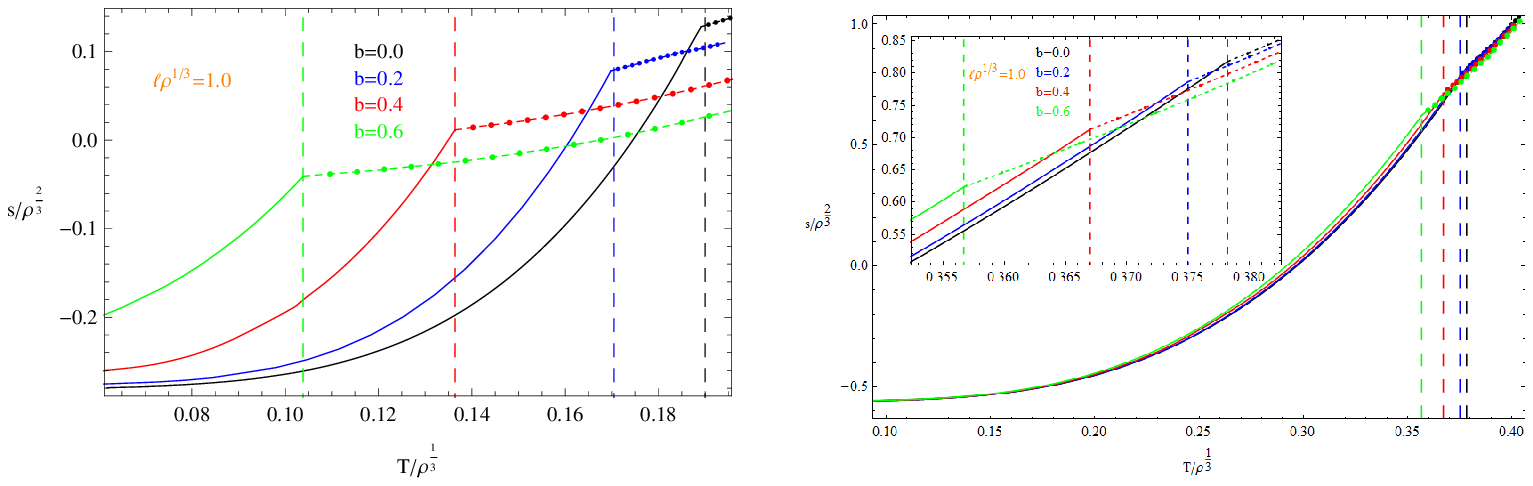}
\caption{\label{d5gzbst}(Color online.) The entanglement entropy of the
operator $\langle O_{+}\rangle$ (left plot) and  the
operator $\langle O_{-}\rangle$ (right plot) with respect to the factor $T$ for
different b as $\ell\rho^{\frac{1}{3}}=1.0$.
The four lines
from bottom to top are for $b=0$ (black), $b=0.2$ (blue), $b=0.4$
(red), and $b=0.6$ (green) respectively.}}

To get further understanding of the effect of the width $\ell$ on the entanglement entropy in the hair phase, we plot the corresponding results in Fig. \ref{d5gfLs} with $b=0.4$.
For the operator $\langle O_{+}\rangle$ presented in the left-hand panel, we find that
with the increase of the factor $\ell$ the value of the entanglement entropy increases.
From the right panel, we can see that the behavior of the entanglement entropy for the  operator $\langle O_{-}\rangle $ with respect of the width $\ell$ is similar to the case of the  operator $\langle O_{+}\rangle$.

\FIGURE{
\includegraphics[scale=0.85]{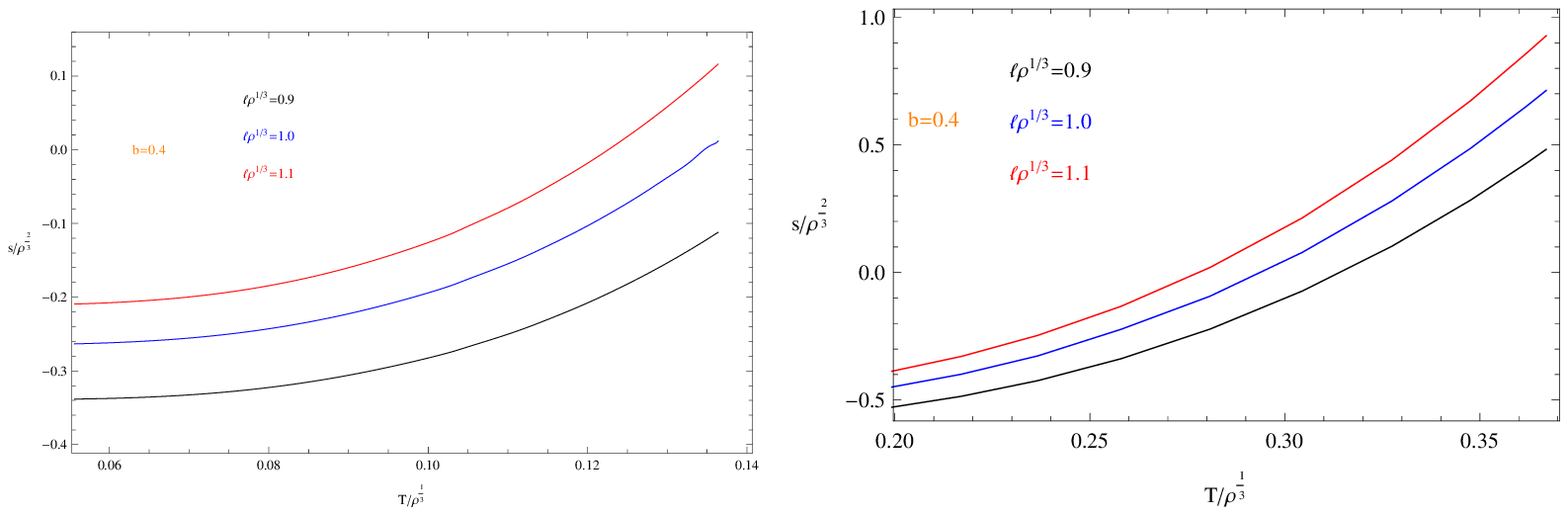}
\caption{\label{d5gfLs}(Color online.) The entanglement entropy of the
operator $\langle O_{+}\rangle$ (left plot) and  the
operator $\langle O_{-}\rangle$ (right plot) with respect to the factor $T$ for
different width $\ell$ as $b=0.4$. The three lines
from bottom to top are for $\ell\sqrt{\rho}=0.9$(black),
$\ell\sqrt{\rho}=1.0$ (blue), $\ell\sqrt{\rho}=1.1$ (red) respectively.}}

\section{Summary}
We studied the phase transition and the behavior of entanglement entropy in AdS black hole gravity with ENE in four and five dimensional spacetimes. Firstly, we revealed the properties of the
phase transition by analyzing the behaviors of the scalar hair. It was found that
the scalar hair emerges at critical temperature $T=T_c$, which implies
the phase transition appears. With the increase of the ENE factor $b$,
the value of the critical temperature $T_c$ becomes smaller. That is to say, the ENE
correction to the usual Maxwell field makes the scalar hair harder
to form in the full-backreaction model.

More importantly, we investigated the holographic entanglement entropy
of the holographic model. For the 4-dimensional spacetime, we observed that the holographic entanglement entropy as a function of temperate has a discontinuous slop at the critical temperature $T_c$. And this jump of the slop of the entanglement entropy
discontinuity corresponds to second order phase transition in the holographic model with exponential nonlinear electrodynamics. Moreover, the entanglement entropy after
condensate is lower than the one in the metal phase and decreases
monotonously as temperature decreases. This is due to the fact that the formation of Cooper pairs make the degrees of freedom decrease in the hair phase. For a fixed T,
we can see that the entanglement entropy of the operator $\langle O_{+}\rangle$ and operator $\langle O_{-}\rangle$ in the normal phase becomes smaller as the ENE factor b increases. In the superconducting phase, however, it is surprising that the influence of the parameter b on the  entanglement entropy of the operator $\langle O_{+}\rangle$ is non-monotonic. We found that for the both  operators the entanglement entropy increases with the increase of the width of the region.

We also explored the holographic entanglement entropy of the holographic model in 5-dimensional spacetime and observed that the behave of the entanglement entropy as a function of the temperature is similar to the case of the 4-dimensional spacetime. Furthermore, it was of interesting to note that the slope of the curve decreases as the temperature is lowered in superconductor situation. For a given T, we found that the entanglement entropy of two operators decreases in the metal phase but increases in the hair phase as we choose a larger ENE factor. We also observed that the effect of the ENE factor on the entanglement entropy is getting smaller as we choose a low temperature in the superconducting phase, and the value of the entanglement entropy becomes bigger as the width of the region increases.

\begin{acknowledgments}
This work was supported by the National Natural Science Foundation of China under Grant No. 11475061; and Construct Program of the National Key Discipline. The talent recruitment program of Liupanshui normal university under Grant No. LPSSYKYJJ201509.
\end{acknowledgments}

\end{document}